\documentclass[%
    article,
    twocolumn,
preprintnumbers,
bibnotes,amsmath,amssymb,
aps,
pre,
floatfix,
]{revtex4-2}

\usepackage{graphicx}
\usepackage{dcolumn}
\usepackage{pifont}
\usepackage{tikz}
\usepackage{circuitikz}
\usepackage{mathtools}
\usepackage{bm}
\usepackage{amsmath,amssymb}
\usepackage{hyperref}
\usepackage{mathtools}
\usepackage{lipsum}
\usepackage{float}
\usepackage{booktabs}
\usepackage[normalem]{ulem}

\newcommand{\mrm}[1]{\mathrm{#1}}

\newcommand{\figRef}[2]{Fig.$\;$#1(#2)}
\newcommand{\figRefWithSubref}[3]{Fig.$\;$#1(#2)(#3)}
\newcommand{\figRefWithTime}[3]{Fig.$\;$#1(#2)$\;$at$\; t=t_{#3}$}

\newcommand{\tX}{\mathtt{X}}
\newcommand{\tY}{\mathtt{Y}}

\newcommand{\arxiv}[1]{\href{http://arxiv.org/abs/#1}{\texttt{arXiv}:#1}}

\begin{document}

\preprint{\arxiv{XXXX.XXXXX}}

\title{\texorpdfstring{Metastable Dynamical Computing with Energy Landscapes: A Primer}{Metastable Dynamical Computing with Energy Landscapes: A Primer}}

\author{Christian Z. Pratt}%
 \email{czpratt@ucdavis.edu}

 \author{Kyle J. Ray}%
 \email{kjray@ucdavis.edu}
 
\author{James P. Crutchfield}
    \email{chaos@ucdavis.edu}
\affiliation{%
Complexity Sciences Center and Department of Physics and Astronomy, University of California, Davis, One Shields Avenue, Davis, CA 95616
}%

\date{\today}

\begin{abstract}

Smartphones, laptops, and data centers are CMOS-based technologies that ushered our world into the information age of the $21^\text{st}$ century. Despite their advantages for scalable computing, their implementations come with surprisingly large energetic costs. This challenge has revitalized scientific and engineering interest in energy-efficient information-processing designs. One current paradigm---dynamical computing---controls the location and shape of minima in potential energy landscapes that are connected to a thermal environment. The landscape supports distinguishable metastable energy minima that serve as a system's mesoscopic memory states. Information is represented by microstate distributions. Dynamically manipulating the memory states then corresponds to information processing. This framing provides a natural description of the associated thermodynamic transformations and required resources. Appealing to bifurcation theory, a computational protocol in the metastable regime can be analyzed by tracking the evolution of fixed points in the state space. We illustrate the paradigm's capabilities by performing $1$-bit and $2$-bit computations with double-well and  quadruple-well potentials, respectively. These illustrate how dynamical computing can serve as a basis for designing universal logic gates and investigating their out-of-equilibrium thermodynamic performance. 
\end{abstract}
\maketitle

\textbf{We review a dynamical computing paradigm that manipulates microstate distributions in a metastable potential energy landscape immersed in a thermal environment. A candidate landscape contains energy minima separated by sufficiently large barriers that stabilize mesoscopic memory states. Changing the landscape’s structure corresponds to information processing. Tracking fixed-point bifurcations then provides a path to designing energy-efficient computations.}

\section{Introduction}
\label{sec:Introduction}

Over recent decades, integrated-circuit technologies contributed to a global abundance of computing---from laptops and smartphones to acre-sized data centers. While potentially offering boundless opportunities for humanity, these devices generate $\mathcal{O}(10^4)$ times more heat \cite{Freitas_Delvenne_Esposito_2021, Gao_Limmer_2021, Takeuchi_2022, Intel} than what is theoretically required \cite{Landauer_1961}. Even before the recent rapid appearance of large language models \cite{kamatala2025transformers}, energy consumption for computational purposes was projected to reach 20\% of global energy demand by 2030 \cite{Jones_2018}. It is clear that these CMOS-based technologies cannot serve global computational needs much longer.

To address the ever-growing demands, alternative computing paradigms are being explored in  industry and academia. One approach, first explicated by Landauer in 1961 \cite{Landauer_1961} and subsequently followed up by others \cite{Bennett_1982, Sekimoto_1997, Plenio_Vitelli_2001, Blickle_Speck_Helden_Seifert_Bechinger_2006, Jun_Gavrilov_Bechhoefer_2014, Riechers_2019, Riechers_Boyd_Wimsatt_Crutchfield_2020, Boyd_Patra_Jarzynski_Crutchfield_2022}, focuses on controlling the minima in the potential energy landscape of a system coupled to a thermal environment. Coarse graining the continuous microstate phase space into a discrete set of high-probability regions surrounding the energy minima yields long-lived mesoscopic metastable memory states. If the landscape contains pairs of energy minima, then these information-bearing degrees of freedom correspond to the logical $0$s and $1$s in binary computing. Changing the memory states' dynamics in the landscape corresponds to performing information processing---creating, destroying, and transforming the $0$s and $1$s. Equivalently, tracking the fixed points of the landscape's associated dynamics permits the design of computational protocols.

We first detail the intuition underlying dynamical landscape computing in Section \ref{sec:FoundationsOfDynamicalLandscapeComputing}. After this, we illustrate the design paradigm by reviewing two computations: Section \ref{sec:OneBitInformationErasure} reviews the $1$-bit information erasure operation \cite{Landauer_1961}; then, Section \ref{sec:TwoBitDynamicalComputations} outlines $2$-bit \emph{control erasure} \cite{Pratt_Ray_Crutchfield_2024}, a generalization of Landauer's erasure in a two-dimensional landscape. These examples demonstrate how dynamical landscapes can serve as a platform both for designing classical logic gates \cite{Pratt_Ray_Crutchfield_2024, ray2023gigahertz} and for investigating their out-of-equilibrium thermodynamic performance \cite{Landauer_1961, Bennett_1982, Jun_Gavrilov_Bechhoefer_2014, Boyd_Patra_Jarzynski_Crutchfield_2022}.

\section{Foundations of Dynamical Landscape Computing}\label{sec:FoundationsOfDynamicalLandscapeComputing}

How can a noisy dynamical system perform classical computations? We will investigate this by building up the canonical energy landscape and its associated dynamics for $1$-bit computing. First, let's consider an arbitrary one-dimensional flat landscape $U(x)$ that is initially unconnected to a thermal environment---see \figRef{1}{a}. We then introduce a particle into the potential to represent system information. In this setup, the particle freely travels in either direction along the $x$-axis according to its initial position and velocity. In other words, there is no notion of discrete information storage.

\begin{figure}[!t]
        \centering
        \includegraphics[width=0.85\linewidth]{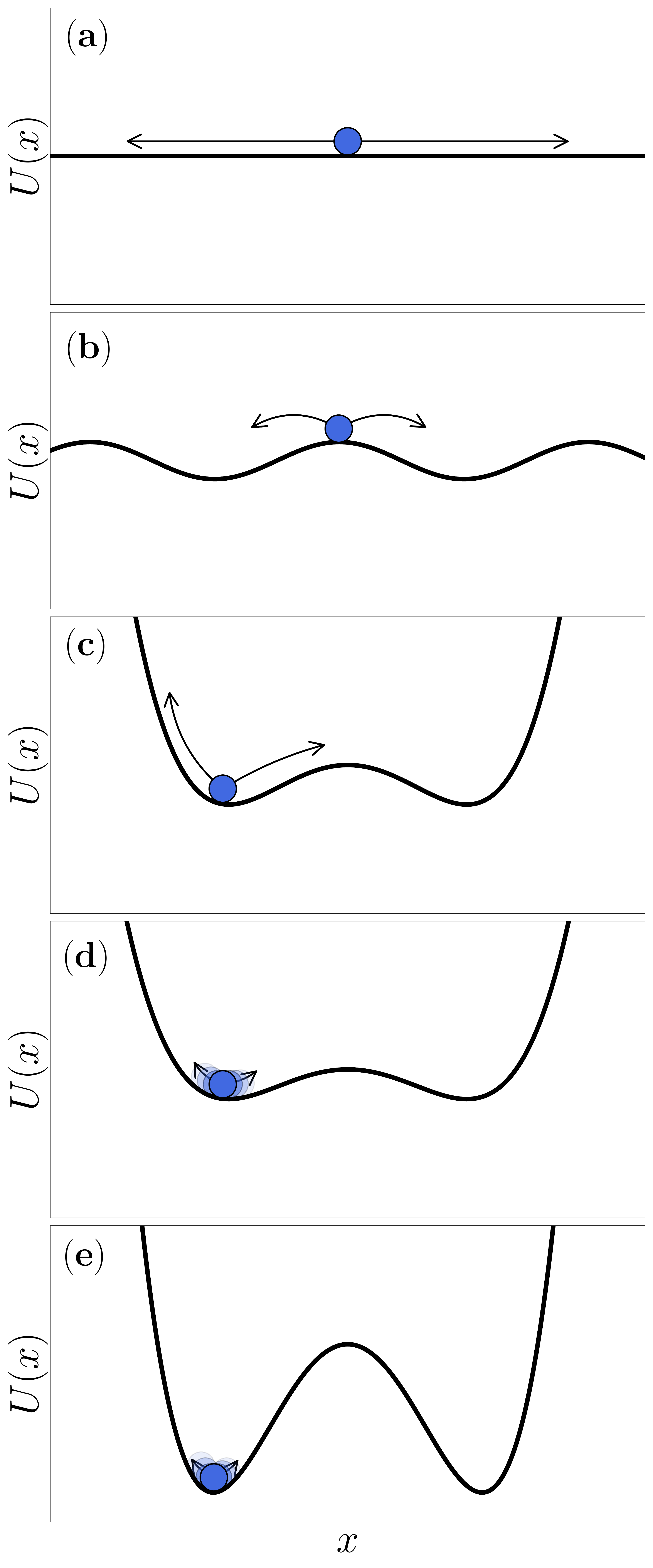}
        \caption{Building a potential energy landscape $U(x)$ that stores and processes a single bit of information. (a) Flat landscape. (b) Landscape with an arbitrary number of energy minima. (c) Instantiating two minima in the landscape, but with no thermal environment connection. (d) Two wells but with damped and noise-perturbed particle motion. (e) Double-well potential with a sufficiently large barrier height for stable information storage.} 
        \label{fig:landscape buildup}
\end{figure}

One way to implement discrete information storage in the landscape is to add energy wells, which are individually separated by energy barriers of equal height. \figRef{1}{b} shows one example, where infinitely many minima are contained in the landscape. Now, coarse grain the continuum of microstates, grouping those into initial positions in each well and total energy less than the barrier height into a discrete information-bearing memory state. Although these wells permit information storage, there are an infinite number of them. In contrast, today's CMOS devices store bits---a unit of information with only two states, $0$ or $1$---to represent information. 

With binary computational states in mind, we simplify the landscape to two wells separated by one energy barrier, as shown in \figRef{1}{c}. These result is binary information storage. The only force acting on the particle comes from the potential $-\mathrm{d}U(x)/\mathrm{d}x$. In other words, the system is conservative: If the particle has low enough energy, it will be confined to its current well; conversely, if it has enough energy to travel over the barrier from one minimum to the other, the particle will continue to oscillate back and forth without losing energy, being unable to settle into either well.

To mitigate this, we couple the system to a thermal environment, i.e., a sufficiently large ideal heat bath at equilibrium temperature $T$. As a result, a damping force $-\gamma \mathrm{d}x/\mathrm{d}t$ with coefficient $\gamma$, as well as a thermal noise term $\eta(t)$, now acts on the particle. A common choice for modeling $\eta(t)$ is Gaussian white noise, such that the particle evolves according to Langevin dynamics \cite{Han_Lapointe_Lukens_1992, Sekimoto_1998}:
\begin{equation}\label{eq:Langevin}
    m\dfrac{\mathrm{d}^2x}{\mathrm{d}t^2} = 
    -\gamma\dfrac{\mathrm{d}x}{\mathrm{d}t} -  \dfrac{\mathrm{d}U(x)}{\mathrm{d}x}  + \eta(t)
    ~.
\end{equation}

Alternatively said, the particle now evolves according to Newton's $2$nd law due to a dissipative force, a conservative force, and a stochastic force. Consequently, the particle favors settling into the phase space region surrounding a minimum, as seen in \figRef{1}{d}. By coupling the system to the heat reservoir, its state is now better represented as a distribution of microstates. The particle can now be interpreted as the mean of this distribution. Thermal noise will randomly perturb the particle within its respective well, mixing and merging trajectories with different histories into the same long-time behavior.

However, rare noise fluctuations can be large enough to cause transitions between the wells. This renders the landscape's mesoscopic memory states to not be purely stable, but \textit{metastable} \cite{Hanggi_1986} as they balance noise and dissipation. For the one-dimensional potential in \figRef{1}{d}, the particle's average energy in a metastable state is on the scale of $k_B T$, where $k_B$ is the Boltzmann constant \cite{Hanggi_1986, Han_Lapointe_Lukens_1989, Han_Lapointe_Lukens_1992}. If the energy barrier height is not much larger than $k_B T$, there is a high probability that the particle will fluctuate over the barrier, producing a thermally-activated information storage error \cite{Han_Lapointe_Lukens_1989, Han_Lapointe_Lukens_1992}. More precisely, the probability of an energy fluctuation of size $\Delta E$ is proportional to $\mathrm{exp}(-\Delta E / k_B T)$ \cite{Hanggi_1986, Han_Lapointe_Lukens_1989, Han_Lapointe_Lukens_1992}, where $\Delta E$ represents the height of the energy barrier.

Reducing these errors involves raising the energy barrier height to be sufficiently higher than $k_B T$, as illustrated in \figRef{1}{e}. Since the distribution is only stable relative to a timescale that depends on the height of the energy barrier, we have now ensured that the state can be stored over long time scales with an exponentially small probability of error. We now deem the phase space regions surrounding minima to be the metastable memory states of the landscape. Together, the mesoscopic states on either side of the $x$-axis represent one bit of information. This landscape serves as our computational substrate for executing $1$-bit computations. 

The double-well potential in \figRef{1}{e} can be described by a 4th-order polynomial:
\begin{equation}\label{eq: 1 bit general potential}
    U(x) = \dfrac{1}{4}ax^4 - \dfrac{1}{2} b x^2 + cx ~,
\end{equation}
where $a$, $b$, and $c$ are real-valued scalars that serve as the potential's control parameters. Generally speaking, this potential can be continuously transformed into a single well depending on the values of $a$ and $b$, while $c$ tilts the potential to the positive (negative) side of the $x$-axis if its value is negative (positive).

To illustrate the role of the associated state-space geometry and the use of bifurcation theory, consider a concrete example of a fixed-point analysis for Eq. \eqref{eq: 1 bit general potential} which can produce \figRef{1}{e}. First, let $a = 1$ so that $b$ provides complete control of the barrier height. Setting $c = 0$ enforces a potential symmetric about $x=0$. To start the fixed point analysis, we find values of $x$ that set the derivative of $U(x)$ with respect to $x$ to zero. These critical values $\{x^*\}$ are the potential's fixed points:
\begin{align}
    \dfrac{\mrm{d}U(x)}{\mrm{d}x} \Big|_{x=x^*} &= x^{*3} - bx^* = 0 ~,
\end{align}
with solutions $x^* = 0, \pm \sqrt{b}$. We have 3 real fixed points when $b$ is positive. We can assess the stability of $U(x=x^*)$ to understand how microstate distributions interact with the potential by evaluating $-\mathrm{d}^2U(x=x^*)/\mathrm{d}x^{*2}$:
\begin{align}\label{eq:1 bit curvature}
    -\dfrac{\mrm{d}^2U(x=x^*)}{\mrm{d}x^{*2}} &= 3x^{*2} - b
    ~.
\end{align}

Next, $x^* = \pm \sqrt{b}$ results in $\mathrm{d}^2U(x^*=\pm \sqrt{b})/\mathrm{d}x^{*2} = 2b > 0$. Now, both fixed points are stable. This validates our choice of the potential in \figRef{1}{e}.

\section{Single-bit Erasure}
\label{sec:OneBitInformationErasure}

By changing the landscape's barrier height and the locations of its minima, particles can be controlled to travel between the system's memory states. Said differently, dynamically manipulating the landscape's metastable states corresponds to information processing. 

We will demonstrate this with two information erasure protocols that are inspired by the the Restore-to-One (RT1) protocol \cite{Landauer_1961}, whose differences can be seen through a bifurcation theory perspective: A pitchfork bifurcation protocol versus a saddle-node bifurcation protocol.

\subsection{Pitchfork protocol}\label{sec:PitchforkBifurcationProtocol}

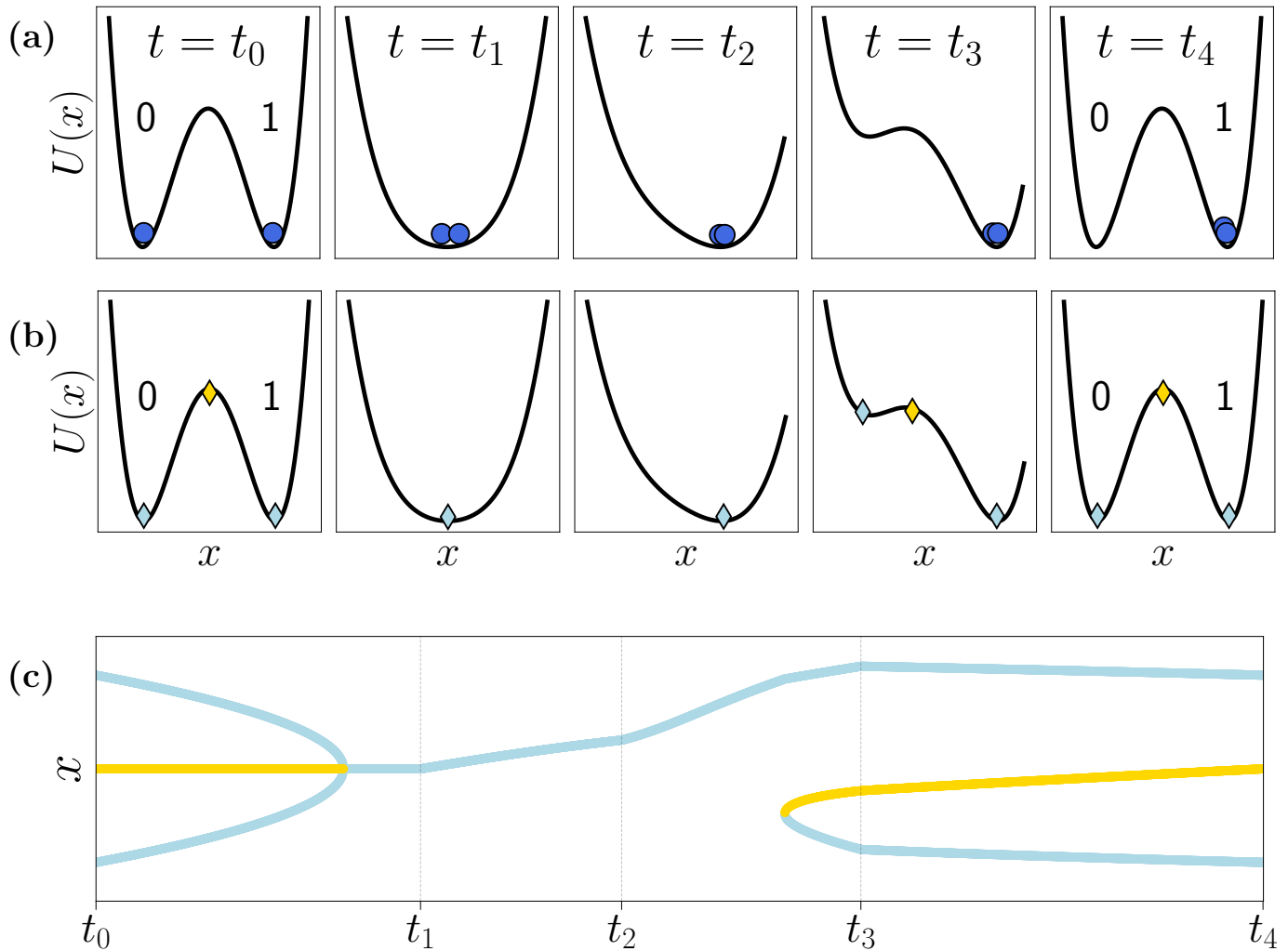
\begin{figure*}[t]
    \centering
        \centering
        \begin{tikzpicture}
            
        \node[inner sep=0pt] (ce) at (0,0){\includegraphics[width=\linewidth]{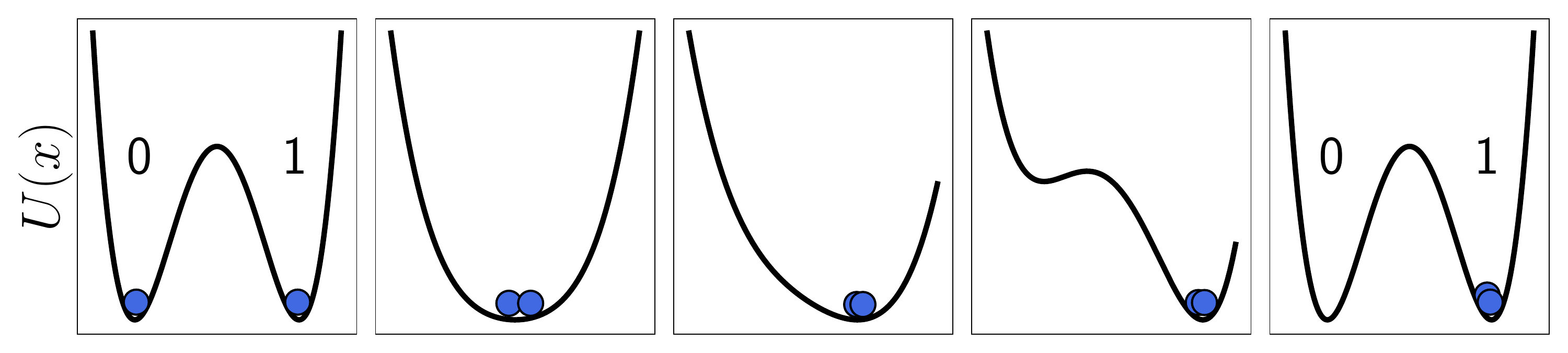}};
        \node[align=center] at (-9, 1.35) {\Large{\textbf{(a)}}};

        \node[align=center] at (-6.5, 1.25) {\huge{$t=t_0$}};

        \node[align=center] at (-3.07, 1.25) {\huge{$t=t_1$}};

        \node[align=center] at (0.5, 1.25) {\huge{$t=t_2$}};

        \node[align=center] at (3.75, 1.25) {\huge{$t=t_3$}};

        \node[align=center] at (7.07, 1.25) {\huge{$t=t_4$}};
        \end{tikzpicture}
        \centering
        \begin{tikzpicture}
            \node[inner sep=0pt] (ce) at (0,0){\includegraphics[width=\linewidth]{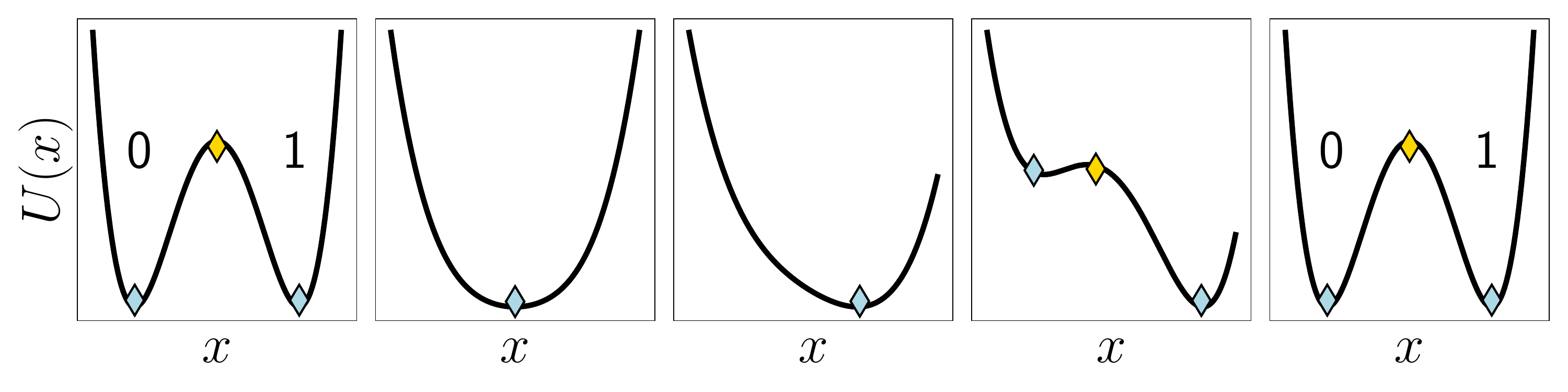}};
        \node[align=center] at (-9, 1.35) {\Large{\textbf{(b)}}};
        \end{tikzpicture}

    \vspace{0.5cm}
    
        \centering
        \begin{tikzpicture}
            \node[inner sep=0pt] (ce) at (0,0){\includegraphics[width=\linewidth]{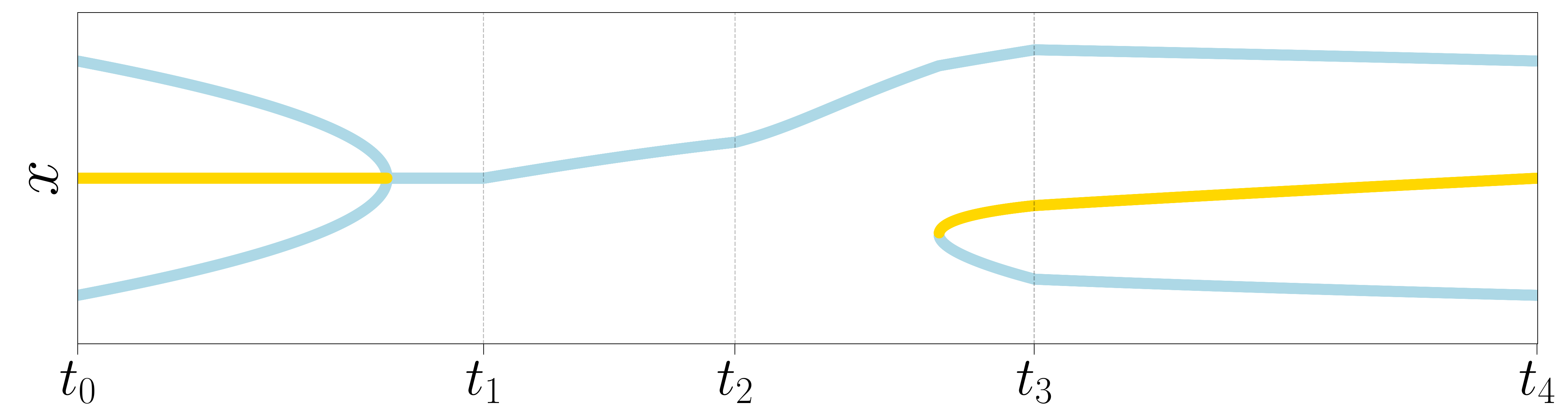}};
            \node[align=center] at (-9, 1.65) {\Large{\textbf{(c)}}};
        \end{tikzpicture}
\caption{A pitchfork information erasure protocol inspired by the Restore-to-One (RT1) protocol \cite{Landauer_1961}
from different dynamical perspectives, taken at different points in time. (a) RT1 protocol showing both possible initial particle locations that end in the $1$ state. (b) RT1 protocol only illustrating the potential's fixed points. (c) A dynamical view of information erasure, demonstrating that erasure protocols are easily understood using bifurcation theory. The blue (yellow) lines represent the stable (unstable) fixed point trajectories throughout the protocol. In the dynamical-systems language overall the protocol is quite simple---a reverse pitch-fork bifurcation followed by a reverse saddle-node bifurcation.}
\label{fig:PitchforkProtocol}
\end{figure*}

We first introduce a particle into the landscape. If is sampled in equilibrium, it has a $50\%$ probability of being in one or the other well. Regardless of the particle's initial well, the goal of the computation is to bring it to the $1$ state \cite{Landauer_1961}. We show this in \figRefWithTime{\ref{fig:PitchforkProtocol}}{a}{0} with each particle representing a potential starting location. 

Assuming that a particle remains in local metastable equilibrium throughout the protocol, we can equivalently view the system response from a bifurcation theory perspective by tracking only the potential's fixed points. A particle then represents a sampling of a locally-stable distribution in the landscape. Its mean location corresponds to a respective landscape's stable fixed point. And, its higher moments are determined by the local curvature of the potential that guides the particle's motion in the region surrounding an energy minima. This perspective is illustrated in \figRefWithTime{\ref{fig:PitchforkProtocol}}{b}{0}, where the potential has two (one) stable (unstable) fixed points that are colored as blue (yellow) diamonds.

First, we lower the energy barrier separating the two wells. This transforms the double-well potential into a single well and causes the two possible particle locations to converge to a single minimum, as displayed in \figRefWithTime{\ref{fig:PitchforkProtocol}}{a}{1}. In dynamical-systems language, this corresponds to a pitchfork bifurcation that forms one single stable fixed point in \figRefWithTime{\ref{fig:PitchforkProtocol}}{b}{1}. We choose the pitchfork bifurcation since we are able to perform this step adiabatically in the quasi-static limit, where each particle remains close to its respective potential energy minima. In this limit, a particle stays arbitrarily close to local equilibrium if the protocol is done arbitrarily slowly. The next subsection showcases a protocol that fundamentally differs in this step and discusses the relative energetic efficiencies.

Next, to ensure that the particle ends in the $1$ state regardless of its initial start state, the potential tilts towards the positive $x$-axis. \figRefWithTime{\ref{fig:PitchforkProtocol}}{a}{2} demonstrates this, where the particle moves into the right half-plane. Correspondingly, the landscape's single stable fixed point is translated towards the positive $x$-axis, as illustrated in \figRefWithTime{\ref{fig:PitchforkProtocol}}{b}{2}.

Now, to maintain the particle's location while completing the computation, we raise the energy barrier to its original height while keeping the potential tilted, as seen in \figRefWithTime{\ref{fig:PitchforkProtocol}}{a}{3}. This recreates the well in the left half-plane. Due to the large barrier height, the particle is likely to stay in right half-plane. This protocol step can be viewed as the creation of one stable and one unstable fixed point, bringing the total number of fixed points to match the start of the protocol as illustrated in \figRefWithTime{\ref{fig:PitchforkProtocol}}{b}{3}. This is a reverse saddle-node bifurcation.

The final step returns the landscape back to its original configuration, completing the protocol cycle. The potential first untilts to bring the landscape back to its initial configuration. In this way and independent of initial location, particles now reside in the $1$ state as desired; see \figRefWithTime{\ref{fig:PitchforkProtocol}}{a}{4}. In \figRefWithTime{\ref{fig:PitchforkProtocol}}{b}{4}, this corresponds to the fixed points converging to their original locations. This final step completes the RT1 protocol. Note that the protocol's cyclicity ensures accounting for all relevant information and thermodynamic costs related to the operation---analogous to forming a closed engine cycle in classical thermodynamics. 

Introducing the bifurcation theory perspective of the RT1 protocol emphasizes that we need to track the entire protocol only from the fixed point paths in the state space. Using this vantage point, we are able to further simplify our understanding of what a computation is---we no longer need to track the full trajectory of the microstate distribution. \figRef{\ref{fig:PitchforkProtocol}}{c} illustrates the entire protocol in terms of the trajectories of the potential's fixed points. 

\subsection{Saddle-node protocol}\label{sec:SaddleNodeProtocol}

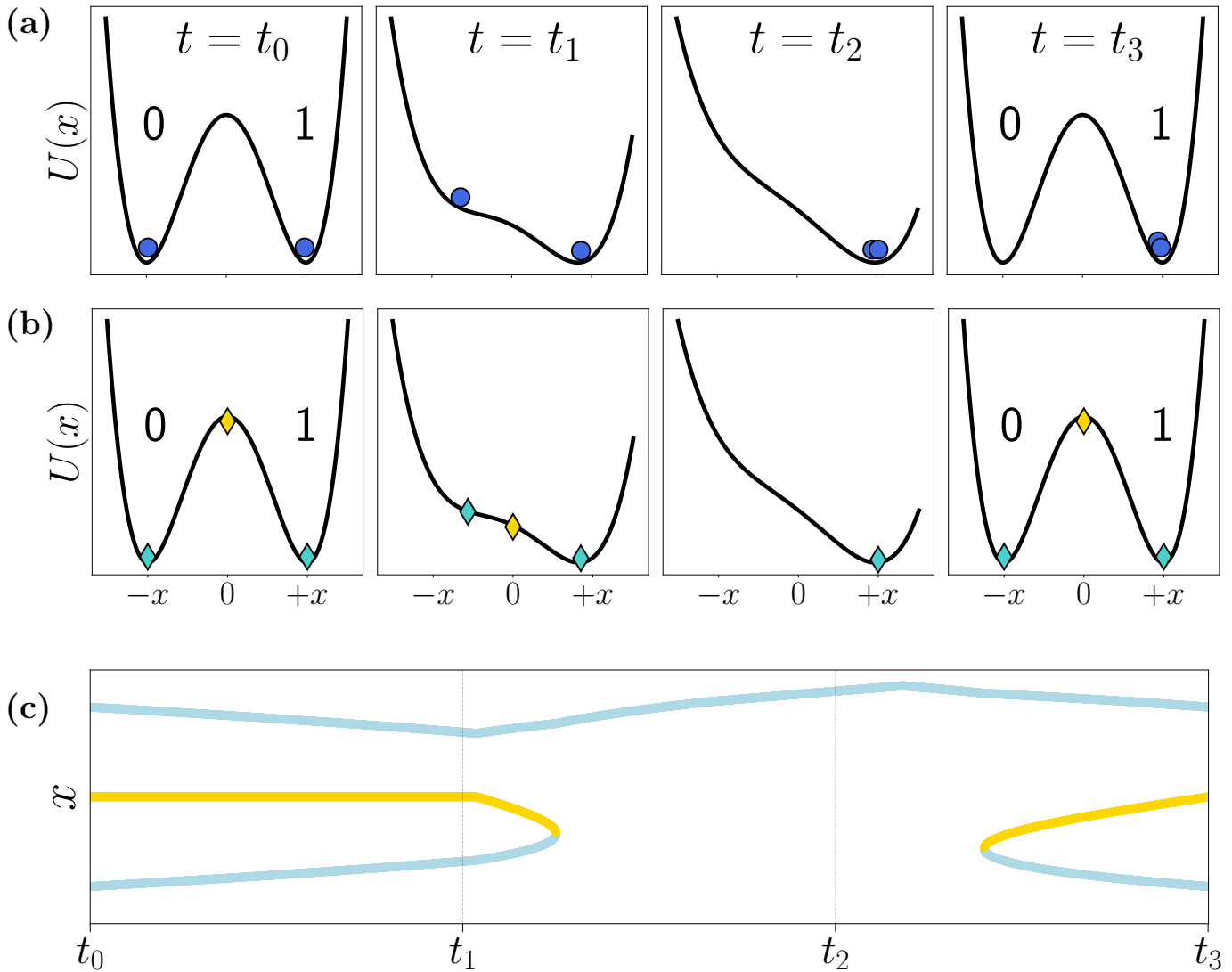
\begin{figure*}[t]
    \centering
        \begin{tikzpicture}
            
        \node[inner sep=0pt] (ce) at (0,0){\includegraphics[width=\linewidth]{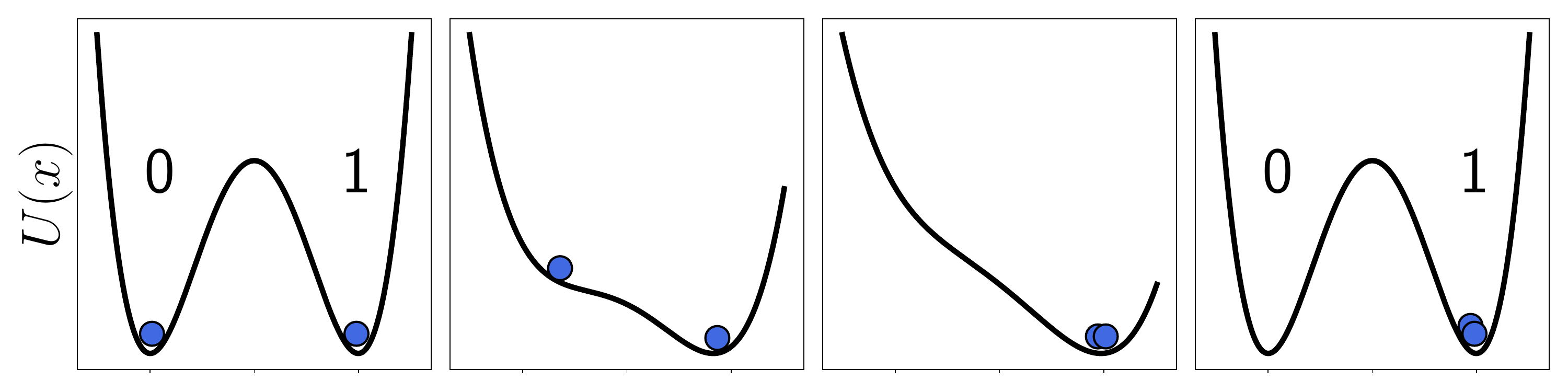}};
        \node[align=center] at (-9, 1.75) {\Large{\textbf{(a)}}};

        \node[align=center] at (-5.95, 1.5) {\huge{$t=t_0$}};

        \node[align=center] at (-1.6, 1.5) {\huge{$t=t_1$}};

        \node[align=center] at (2.6, 1.5) {\huge{$t=t_2$}};

        \node[align=center] at (6.8, 1.5) {\huge{$t=t_3$}};

        \end{tikzpicture}
        \centering
        \begin{tikzpicture}
            \node[inner sep=0pt] (ce) at (0,0){\includegraphics[width=\linewidth]{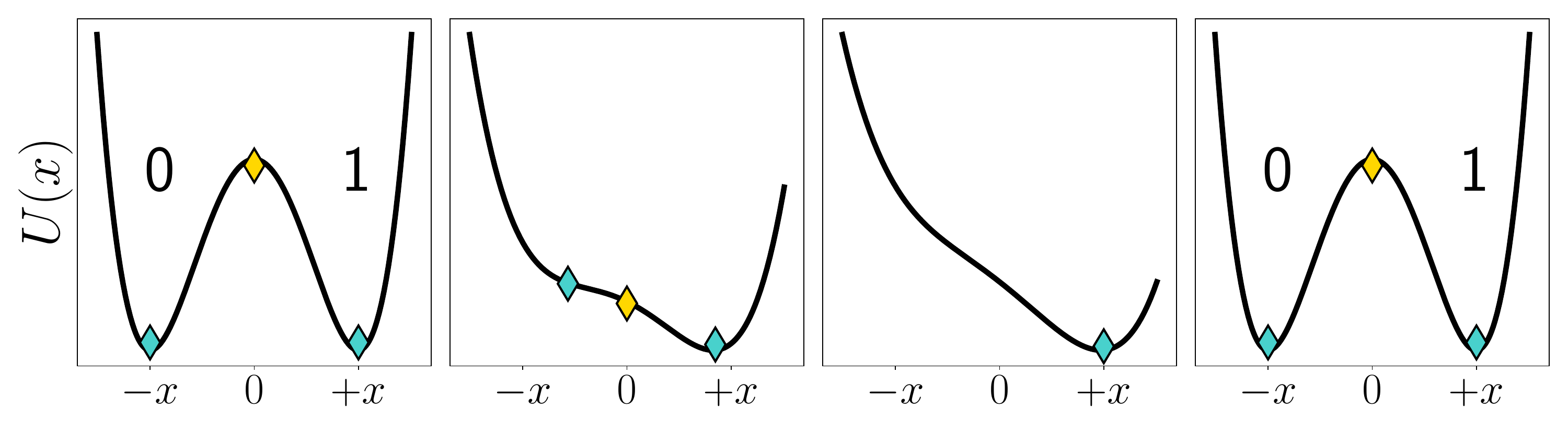}};
        \node[align=center] at (-9, 2) {\Large{\textbf{(b)}}};
        \end{tikzpicture}

    \vspace{0.5cm}
    
        \centering
        \begin{tikzpicture}
            \node[inner sep=0pt] (ce) at (0,0){\includegraphics[width=\linewidth]{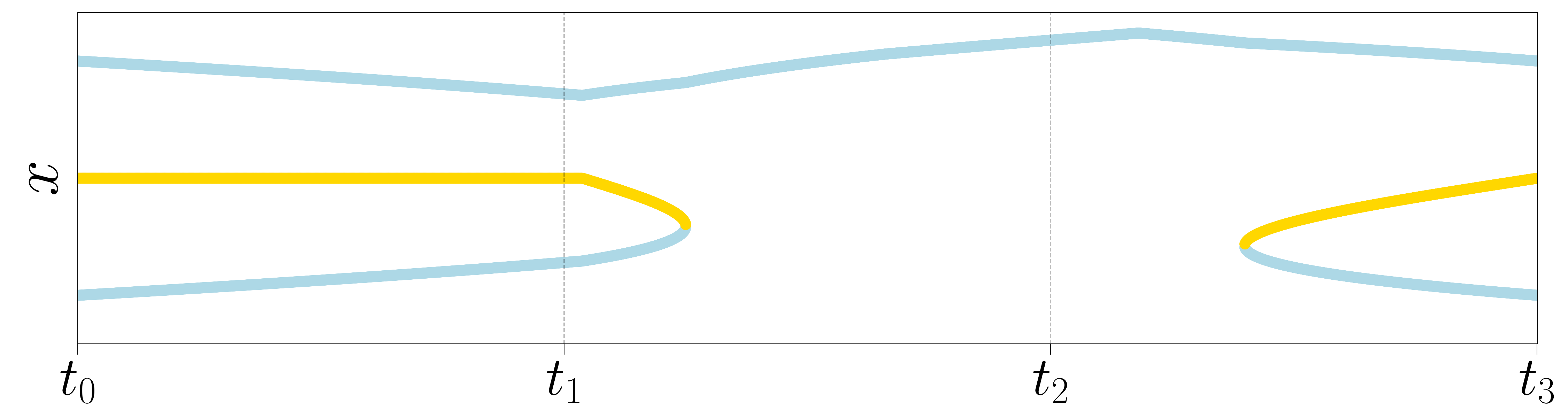}};
            \node[align=center] at (-9, 1.65) {\Large{\textbf{(c)}}};
        \end{tikzpicture}

    \caption{A saddle-node bifurcation information erasure protocol inspired by Ref. \cite{Landauer_1961}. Differing from the pitchfork protocol, it can not be carried out infinitely slowly to make it adiabatic and energy inefficient. Throughout the entire protocol, there are two saddle-node bifurcations. (a) The evolution of the potential during the protocol when keeping track of the microstate distributions. (b) The same protocol, but following the landscape's fixed points. (c) The saddle-node erasure protocol in terms of the landscape's fixed points only.}
    \label{fig:SaddleNodeProtocol}
\end{figure*}

As an alternative to relying on a pitchfork bifurcation, now perform a RT1 protocol that uses a saddle-node bifurcation. Here, the potential's parameters are tuned differently than with pitchfork erasure, but all possible input states are still erased to the $1$ state.

In \figRefWithTime{\ref{fig:SaddleNodeProtocol}}{a}{0}, the potential and particles are initialized in the same manner as the previous protocol. Likewise, \figRefWithTime{\ref{fig:SaddleNodeProtocol}}{b}{0} shows the same starting locations of the fixed points as \figRefWithTime{\ref{fig:PitchforkProtocol}}{b}{0}.

At $t = t_1$, we drop the barrier slightly, as well as tilt the potential towards the positive $x$-axis, as shown in \figRefWithTime{\ref{fig:SaddleNodeProtocol}}{a}{1}. Then \figRefWithTime{\ref{fig:SaddleNodeProtocol}}{a}{2} (\figRefWithTime{\ref{fig:SaddleNodeProtocol}}{b}{2}) shows the completion of this step from a particle (fixed point) perspective. Manipulating the potential in this way annihilates the unstable fixed point with the stable fixed point on the negative $x$-axis, while maintaining the fixed point on the positive $x$-axis. This is the essence of the saddle-node bifurcation: We are able to store information on one side of an axis throughout the protocol, while annihilating other fixed points in the landscape. 

The final step re-raises the energy barrier and un-tilts the potential, as shown in \figRefWithTime{\ref{fig:SaddleNodeProtocol}}{a}{3} and \figRefWithTime{\ref{fig:SaddleNodeProtocol}}{b}{3}. This completes a protocol cycle, returning the potential back to its original configuration and bringing both microstate distributions to the $1$ state. 

Now, having completed a saddle-node erasure, let's detail the differences between this protocol and the pitchfork erasure. In the saddle-node protocol and due to how the landscape is controlled, the particle stored in the $0$ state dissipates energy as it travels from the bifurcation point where it's local minima vanishes down to the well associated with the $1$ state. Such a saddle-node bifurcation cannot be carried out adiabatically, even in the quasi-static limit. This manifested as more irreversible dissipation into the heat bath, along with the Landauer bound cost---the inevitable work cost needed to perfectly erase an unbiased input bit, that being $k_BT \ln 2$ \cite{Landauer_1961}. 

This fundamentally differs from the pitchfork protocol. There, since the particles remain in the phase space region surrounding their respective local minima throughout the entire protocol, the process can be made adiabatic when performed infinitely slowly, allowing it to reach the Landauer limit. From a dynamical-systems theory perspective, because we form one stable fixed point from the bifurcation of two stable and one unstable fixed point, the pitchfork protocol can be more energy efficient. 

By viewing these computations in terms of the landscape's fixed point dynamics, thermodynamic costs can be more readily understood. Additionally, considering fixed-point creations and annihilations allows for analyzing dynamical computations in higher dimensions, which is the subject of the next section.

\section{2-bit Dynamical Computations}
\label{sec:TwoBitDynamicalComputations}

\begin{figure}[t]
    \centering
        \begin{tikzpicture}
        \node[inner sep=0pt] (ce) at (0,0){\includegraphics[width=\linewidth]{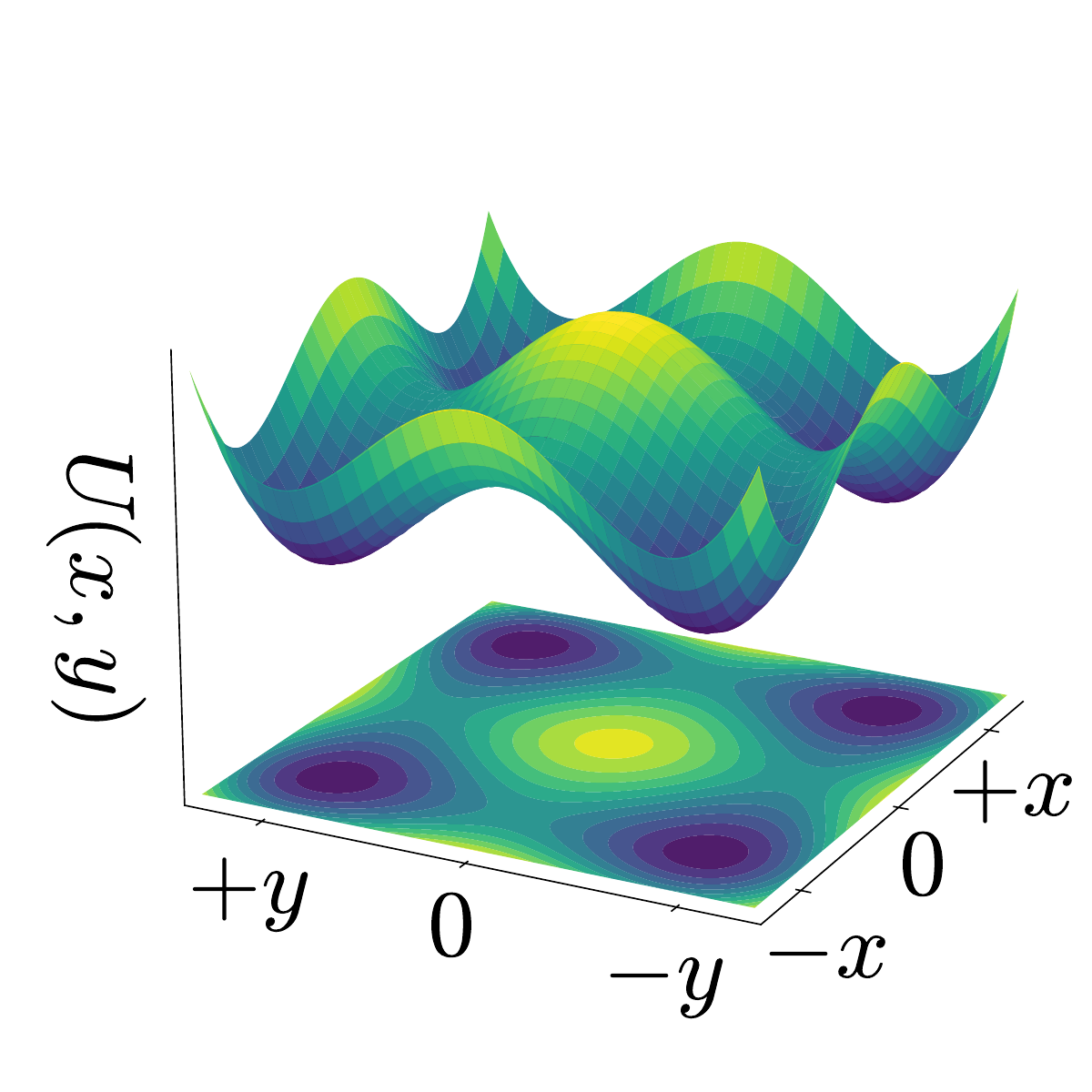}};
        \node[align=center] at (-3.5, 2.25) {\Large{\textbf{(a)}}};
        \end{tikzpicture}
    \centering
        \begin{tikzpicture}
        \node[inner sep=0pt] (ce) at (0,0){\includegraphics[width=\linewidth]{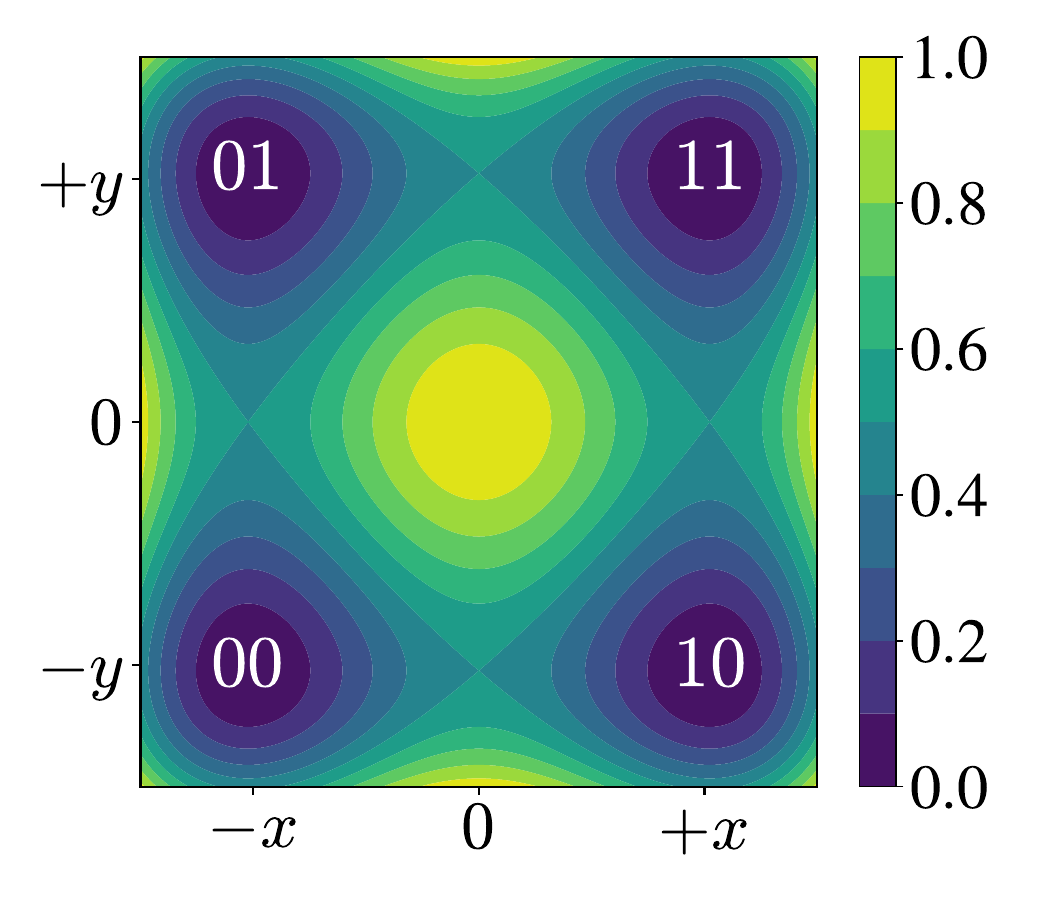}};
        \node[align=center] at (-3.5, 4) {\Large{\textbf{(b)}}};
       \end{tikzpicture}
\caption{Potential energy landscape that supports carrying out $2$-bit computations: (a) Illustrating the two-dimensional landscape $U(x, y)$ via a three-dimensional landscape plot above its projected contour plot. (b) Contour plot of $U(x, y)$ showing example memory-state instantiations.
    }  
\label{fig:quadruple well potential}
\end{figure}

\begin{figure*}[t]
    \centering
        \centering
        \begin{tikzpicture}
        \node[inner sep=0pt] (ce) at (0,0){\includegraphics[width=0.9\linewidth]{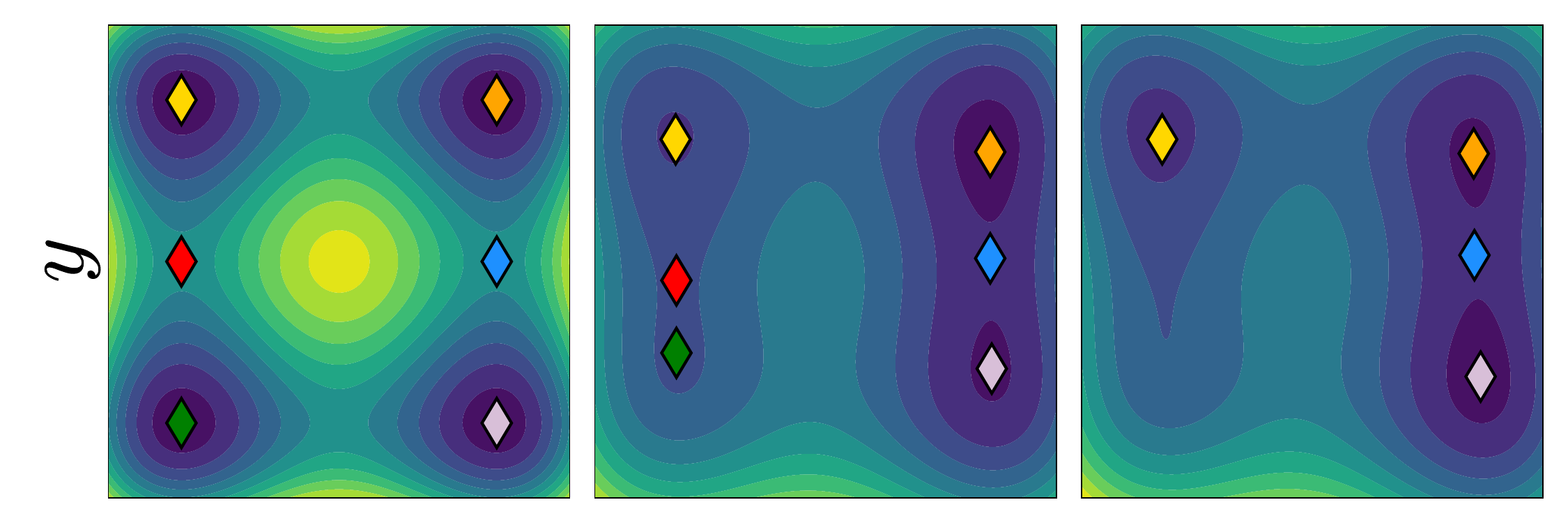}};
        \node[align=center] at (-7.75, 1.7) {\Large{\textbf{(a)}}};
        \node[align=center] at (-4.62, 2.75) {\Large{(i)}};
        \node[align=center] at (0.3, 2.75) {\Large{(ii)}};
        \node[align=center] at (5.30, 2.75) {\Large{(iii)}};
        \end{tikzpicture}
        \centering
        \begin{tikzpicture}
        \node[inner sep=0pt] (ce) at (0,0){\includegraphics[width=0.9\linewidth]{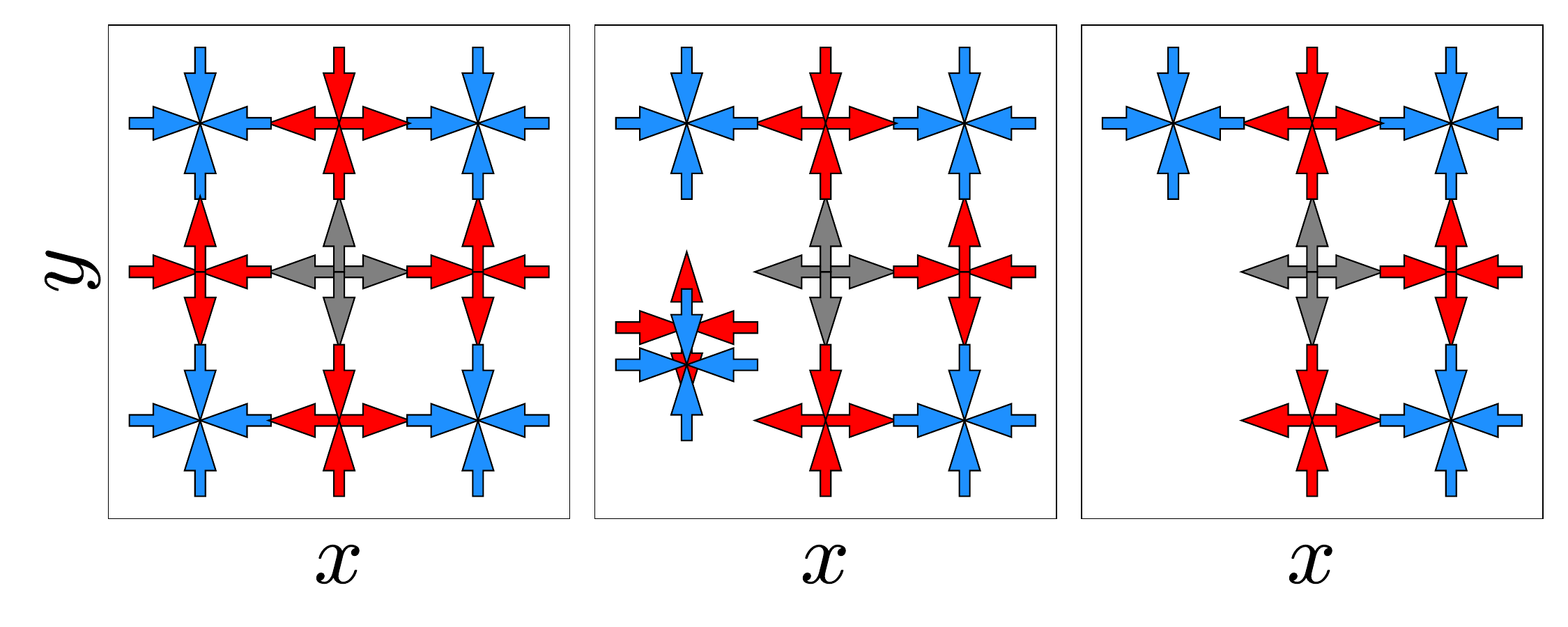}};
        \node[align=center] at (-7.75, 2) {\Large{\textbf{(b)}}};
        \end{tikzpicture}
\caption{Dynamical view of the control erasure (CE) protocol: (a) The CE characterized by the annihilation of the red and green fixed points in the third quadrant, while maintaining all other fixed points. (b) A dynamical skeleton of the CE that only shows the landscape's fixed points and flow fields. Blue (red) corresponds to stable (unstable) fixed points, as well as a grey local maxima. The steps of the protocol are as follows: (i) Initialization. (ii) Drop barrier separating the green (red) stable (unstable) fixed points, while maintaining all other fixed points. (iii) Just after the red-green fixed point annihilation.}
\label{fig:2BitDynamicalComps}
\end{figure*}
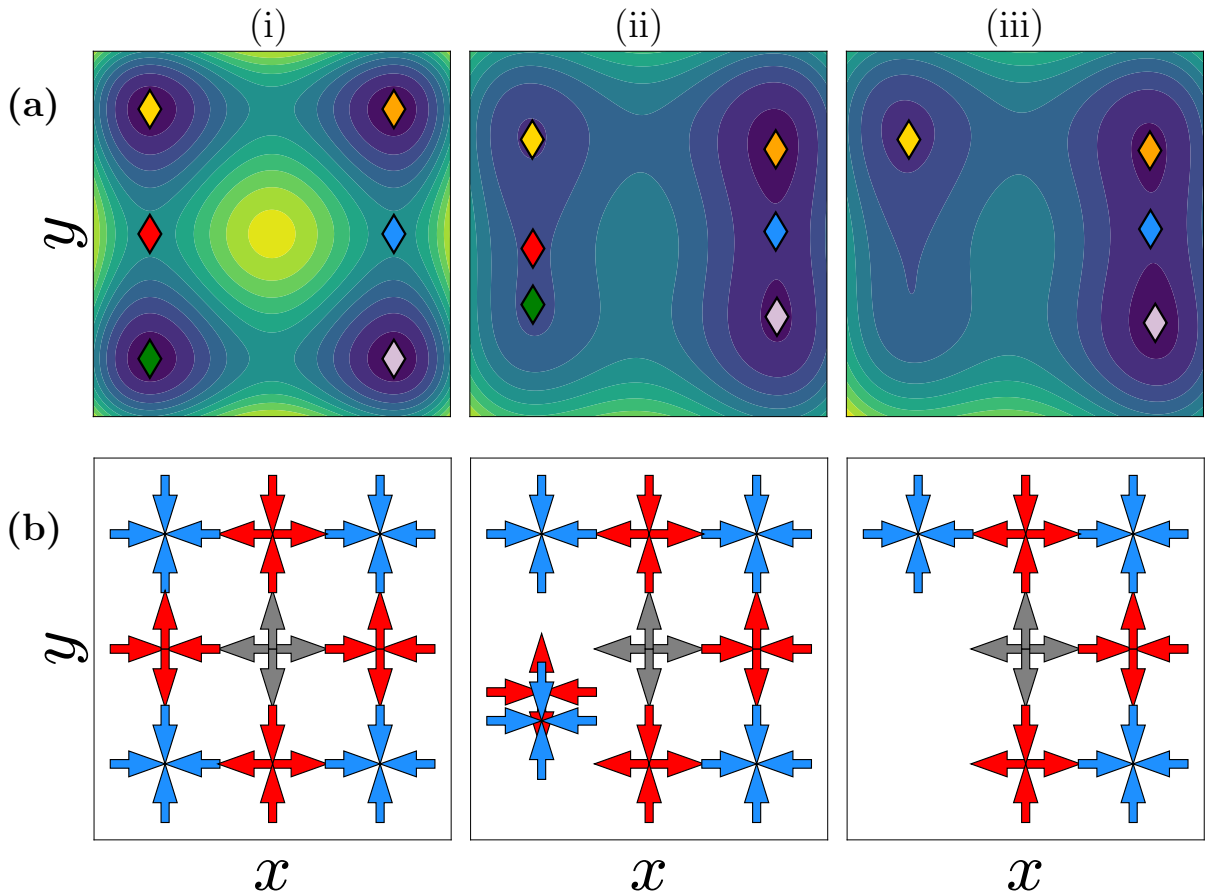

Let's extend the landscape computing paradigm to implement processing two bits of information. Previously, to represent one bit of information required a double-well landscape. This suggests that a landscape that contains four wells provides four different memory states. \figRef{\ref{fig:quadruple well potential}}{a} shows an example potential that contains four wells, each separated by sufficiently large energy barriers, and connected to a thermal environment, as required. Similarly to Eq. \ref{eq: 1 bit general potential}, its form can be created from a general quartic function, but in two dimensions:
\begin{align}\label{eq:quadruple well potential}
    U(x, y) &= \dfrac{1}{4}ax^4 - \dfrac{1}{2}bx^2 + cx \notag \\[1pt]
    & \quad + \dfrac{1}{4}dy^4 - \dfrac{1}{2}ey^2 + fy  \notag \\[1pt]
    & \quad + g\,xy
    ~.
\end{align}
The landscape's control parameters are now $a$, $b$, $c$, $d$, $e$, $f$, and $g$. Allowing $a = d = 1$, and $c = f = g = 0$ produces the example landscape displayed in \figRef{\ref{fig:quadruple well potential}}{a}; changing the values of $b$ and $e$ determines the barrier height along the $y$ and $x$ axes, respectively. Note that if $b \neq e$, then the potential has asymmetrical barrier heights with respect to either axis. 

Performing a bifurcation analysis in a similar way to Sec. \ref{sec:FoundationsOfDynamicalLandscapeComputing} yields four stable fixed points, corresponding to the four wells in the potential; four saddle points, associated with the transitions between adjacent memory states; and an unstable point at the origin. We coarse grain the potential so that its memory states are distinguished according to the minimas' location with respect to the $x$ and $y$ axes:
\begin{align}
\label{eq:mem state instantiation rule}
    \tX \; [\tY] = \begin{cases}
        & 0 \text{\; if \; } x \ [y] < 0 ~, \\
        & 1 \text{\; if \;  }x \ [y] > 0 ~.
    \end{cases} 
\end{align}
This choice of memory state assignment is shown in Fig. \ref{fig:quadruple well potential}(b). To illustrate an example of a computation with this landscape, we first detail the saddle-node erasure protocol that was introduced in Ref. \cite{Pratt_Ray_Crutchfield_2024}. There, a particular physical device was assumed, which provides a potential energy surface that is qualitatively similar to Eq. \eqref{eq:quadruple well potential}, but it has a different mathematical form and control parameters. 

We start by considering the landscape's fixed points, which are depicted as colored diamonds in \figRefWithSubref{\ref{fig:2BitDynamicalComps}}{a}{i}. Further coarse graining, the dynamical point of view directly reveals the potential's fixed points and flow fields, shown in \figRefWithSubref{\ref{fig:2BitDynamicalComps}}{b}{i}. Here, blue (red) arrows represent stable (unstable) fixed points along with a gray maxima. This ``dynamical skeleton'' illustrates where microstate distributions are favored to travel without being concerned about the potential's exact shape. Analogous to Sec. \ref{sec:SaddleNodeProtocol}, we need to lower the barrier separating a minimum, but because we have a four-well potential, we must subsequently tilt the potential in a way that erases some information on one side of an axis, but stores the rest on the other axis. The simplest, and most energy-efficient, option is to implement a pitchfork bifurcation erasure to merge the two wells in the negative $x$ semi-plane. However, such a pitchfork protocol is not possible with the form of the potential in Eq. \ref{eq:quadruple well potential} without destroying stability between the other pair of wells; see Appendix \ref{app:ImpossiblePitchforkWithQuadrupleWellPotential} for the mathematical details.  

Instead, we implement the erasure using a saddle-node bifurcation, as in \figRef{\ref{fig:2BitDynamicalComps}}{a}, by lowering the barrier in the left half-plane and then tilting the potential in such a way to annihilate the red and green fixed points while maintaining all other fixed points. The situations just before and after the annihilation are shown in \figRefWithSubref{\ref{fig:2BitDynamicalComps}}{a}{ii} and \figRefWithSubref{4}{a}{iii}, as well as in \figRefWithSubref{\ref{fig:2BitDynamicalComps}}{b}{ii} and \figRefWithSubref{\ref{fig:2BitDynamicalComps}}{b}{iii}, respectively. Here, the information that was in the third quadrant is erased into the second quadrant, while all other information is preserved. Completing the protocol involves re-raising and un-tilting the potential in an analogous way to the RT1 computation. This erasure protocol is called a \textit{control erasure} (CE) \cite{Pratt_Ray_Crutchfield_2024}: Its functionality involves controlling what input information is erased or preserved in order to perform the $2$-bit computation. Interestingly, there are eight possible CEs that can be performed with the quadruple well potential \cite{Pratt_Ray_Crutchfield_2024}. 

\section{Conclusion}
\label{sec:Conclusion}

With the attenuation of Moore's law colliding with the accelerating demand for computational resources, now is the time to investigate alternative computing paradigms. This is critical for future progress in computing. Here, to address this and show a promising way forward, we described a design strategy for performing computations with dynamical potential-energy landscapes. This perspective of computing brings dynamical-systems theory, statistical mechanics, and nonequilibrium thermodynamics together in a coarse-grained approach to understand thermally-activated classical computations. This computing framework differs from other related dynamical-systems paradigms which assume a physical substrate, such as chaos computing \cite{munakata2002chaos, majumder2018chaos} and nonlinear memristive computing \cite{yang2013memristive, yang2022nonlinearity}. Detailed investigations of thermodynamic quantities and higher-dimensional computations are readily accessible with this framework, opening up new options for optimizing computing systems both in terms of raw performance and energy costs.

\section*{Acknowledgments}

We thank Greg Wimsatt for insightful discussions. The authors also thank the Telluride Science Research Center for its hospitality during visits and the participants of the Information Engines workshop there for their valuable feedback. J.P.C. acknowledges the kind hospitality of the Santa Fe Institute. This material is based on work supported by, or in part by, the U.S. Army Research Laboratory and U.S. Army Research Office under Grant No. W911NF-21-1-0048, and by the Art and Science Laboratory via a gift to UC Davis.

\appendix

\section*{Author Declarations}

The authors have no conflicts to disclose.

\section*{Author Contributions}

All authors developed thermodynamic landscape computing. CZP was responsible for generating the simulation data, formal analysis, and visualizations. KJR and JPC verified the theoretical framework and supervised the development. JPC provided funding. All authors contributed to the manuscript writing and editing and approved the final version for submission.

\section*{Competing Interests}

The authors declare no competing interests.

\section*{Data Availability}

The data that support this review are available from the corresponding author upon reasonable request.

\section{Impossibility of a Pitchfork Bifurcation in a Specific Landscape}\label{app:ImpossiblePitchforkWithQuadrupleWellPotential}

Following Ref. \cite{Wiggens_2003, Strogatz_2018} and as discussed in Section \ref{sec:PitchforkBifurcationProtocol}, a pitchfork bifurcation occurs when two fixed points on either side of an axis converge to a point of annihilation at the coordinate axis. We will now show the impossibility of performing a pitchfork bifurcation on a pair of two adjacent wells in the potential generated by Eq. \eqref{eq:quadruple well potential} and displayed in Figure \ref{fig:quadruple well potential} without destroying stability of the other pair of wells as well. As a consequence, the control erasure (CE) described in Section \ref{sec:TwoBitDynamicalComputations} can only be performed using a saddle-node bifurcation.

First, consider a pitchfork bifurcation in a one-dimensional potential $U(x)$, which is useful for analyzing the two-dimensional case. A common ordinary differential equation that produces a one-dimensional pitchfork bifurcation in the landscape $U(x)$ takes on the polynomial form \cite{Strogatz_2018}:
\begin{equation}
    \dot{x} = -\dfrac{\mrm{d}U(x)}{\mrm{d}x}= rx - x^3 ~. \label{eq:bifurcation ode} 
\end{equation}
Upon observation, the bifurcation parameter appears in the linear term, and Eq. \ref{eq:bifurcation ode} is odd. Said another way, a pitchfork bifurcation is found when the local expansion around a bifurcation point is an odd function with nonvanishing linear and cubic terms \cite{Wiggens_2003, Strogatz_2018}. 

Now, the mathematical form of Eq. \ref{eq:quadruple well potential} does not permit carrying out a CE using a pitchfork bifurcation. To see this, assume that a pitchfork bifurcation can occur at the point $(x,y) = (x_0, y_0)$ in Eq. \ref{eq:quadruple well potential}. Then, expand the first derivative of the potential $U(x,y)$ around $x_0$:
\begin{align}
    & \partial_x U(x_0+h, y_0) \nonumber \\
    & = \partial^2_{x}U(x_0,y_0)h + \partial^3_{x}U(x_0,y_0)h^2/2 + \partial^4_{x}U(x_0,y_0)h^3/6 \label{eq:expansion} \\
    & = (3ax_0^2 - b)h + 3ax_0 h^2 + ah^3 \label{eq:evaluated expansion}
\end{align}
Enforcing that this function takes the form of Eq. \eqref{eq:bifurcation ode} means that the linear and cubic terms remain while the quadratic one vanishes. If $a=0$ then the cubic term vanishes, so we find that $x_0=0$ is the only way for the quadratic term to vanish. Now, because $x_0= 0$, this means that either $b\neq0$ or the linear term vanishes. In order to satisfy the criteria for a pitchfork bifurcation, then $x_0 = 0$ and $a,b \neq0$. Accordingly, we restrict our analysis only to those fixed points at $(x = 0, y = y_0)$. 

As detailed in Sec. \ref{sec:FoundationsOfDynamicalLandscapeComputing}, the stability of a candidate fixed point is controlled by the curvature of the potential at $x= x_0$. This gives $\partial^2_xU(0, y) = -b$. For this to be a stable fixed point, then $b < 0$. That said, suppose this pitchfork bifurcation happens with the $00$ and $01$ wells in the negative $y$ semi-plane from Figure \ref{fig:quadruple well potential}. At the same time, to perform a CE, we must maintain the $10$ and $11$ states, as well as the respective energy barrier separating them. This requires $b > 0$, which contradicts our original restriction of $b < 0$. As a result, the bifurcation cannot happen at $x_0 = 0$, and so it is not a pitchfork bifurcation \cite{Strogatz_2018}. 

In other words, we are unable to form a stable fixed point from an annihilation of the two lower wells, while also maintaining the energy barrier in the upper half-plane. As a result, to perform a CE with the potential in Eq. \ref{eq:quadruple well potential}, we are limited to the less efficient saddle-node bifurcation as discussed in Section \ref{sec:TwoBitDynamicalComputations}.

\bibliography{mainbib}

\end{document}